# Boundary conditions in the envelope function approximation as applied to semiconductor heterostructures: the multi-band case

M.-E. Pistol
Solid State Physics,
Box 118, Lund University
S-221 00 Lund Sweden

Abstract: We have found the equations that determine the self-adjoint extensions, and thus the boundary conditions, of the differential operator used in the multi-band **k·p**-theory, when the coefficients in the Kane-matrix are piecewise constant. Both the one-dimensional and the three-dimensional case have been investigated. A numerical calculation shows that the choice of boundary conditions affects the energy eigenvalues for a quantum well.

**I. Introduction**

The boundary conditions in the envelope function approximation have been studied since a long time [1-2]. It has been shown that the boundary conditions used in the envelope function approximation can include discontinuous wavefunctions in semiconductor heterostructures [2-4]. It was also shown that all symmetric Hamiltonians are equivalent as long as the coefficients are piecewise constant [3] or even piecewise twice differentiable (under mild constraints) [4]. The boundary conditions can be obtained in different ways, for instance based on group theory [5-6] which can give quite general boundary conditions. Other approaches have included the use of Fourier transformation of the multi-band operator which avoids the use of explicit boundary conditions [7]. However, the operator to be transformed can be written in several (actually infinite) symmetric ways giving different transforms. This corresponds to different boundary conditions. The general three-dimensional multi-band case is very complex but has also been considered [8], although not in full generality. In fact the full three-dimensional multi-band case has not been treated in full generality [9] which will be done here.

The most general method to find boundary condition is to use the theory of self-adjoint extensions [10]. Operators in quantum mechanics are usually defined (and self-adjoint) over functions which are twice differentiable everywhere and square integrable. This domain is, however, only a dense subset of the full Hilbert space and if there are points of discontinuities, such as if the effective mass is discontinuous, a larger domain of functions is needed. On this larger domain of functions, which include functions with discontinuities in their values and first derivatives, boundary conditions must be imposed such that the operator remains self-adjoint, i. e. the self-adjoint extensions need to be found. We will use the theory of self-adjoint extensions and find the self-adjoint extensions in the full multi-band case both in one dimension and three dimensions. We will work only in the traditional **k·p**-theory instead of more advanced theories [11-12].

**II. The one-dimensional case**

Let $L$ be a matrix involving differential operators, such as the operator involved in the multi-band **k·p**-approximation. We usually want to solve

$$L \mathbf{u} = E \mathbf{u} \qquad (1)$$

where **u** is a vector in which the components $u_i$ are functions belonging to the Hilbert space $H_i$ and where E is an eigenvalue. The solutions of Eqn. (1) belong to the Hilbert space $H_1 \otimes H_2 \otimes ... \otimes H_n$, i. e. the direct (or tensor) product of the $H_i$'s and n is the order of *L* which is eight in the commonly used eight-band **k·p**-approximation [8,13]. The coefficients of *L* are usually piecewise constant functions of x, where the points of discontinuities are at the interfaces between different semiconductors. The functions $u_i$ are twice differentiable everywhere, except at the points of discontinuities of *L* where they have finite first and second derivatives.

The condition for *L* to be self-adjoint is:

$$\langle L\mathbf{u} \mid \mathbf{v} \rangle - \langle \mathbf{u} \mid L\mathbf{v} \rangle = 0 \tag{2}$$

If we demand that the boundary conditions should be local to each interface we can assume only one interface in the structure (a plane at position x=0) and we have to evaluate Eqn. (2) in the form

$$\int_{-\infty}^{\infty}(L\mathbf{u})^* \cdot \mathbf{v} - \mathbf{u}^* \cdot (L\mathbf{v})dx = \int_{-\infty}^{\infty} \langle L\mathbf{u},\mathbf{v}\rangle - \langle \mathbf{u}, L\mathbf{v}\rangle dx \tag{3}$$

where the * indicates complex conjugation and a transpose of the vector or matrix, i. e. the adjoint operation. We will often use the notation $\langle \mathbf{u},\mathbf{v} \rangle$ instead of $\mathbf{u}^* \cdot \mathbf{v}$ due to its similarity to a scalar product (despite being a function). It is at this stage convenient to express *L* as a sum of three operators

$$L = \mathbf{V} + \mathbf{F}\,\partial_x + \mathbf{S}\,\partial_x^2 \tag{4}$$

where **V**, **F**, and **S** do not contain derivatives, and where, for example, $\mathbf{F}\,\partial_x$ means that every element in **F** is multiplied by $\partial_x$. We observe that an expression like $\mathbf{F}\,\partial_x\,\mathbf{u}$ is associative and that **F** and $\partial_x$ commute. By inspection of the **k·p**-matrix given by Kane [13], and for convenience also given in the appendix, we find that the adjoint matrices are given by

$$\begin{aligned}\mathbf{V}^* &= \mathbf{V}\\ \mathbf{F}^* &= -\mathbf{F}\\ \mathbf{S}^* &= \mathbf{S}\end{aligned} \tag{5}$$

Eqns. (5) are only true if the replacement $k_x \to \partial_x/i$ is done in the **k·p** matrix. We note here that

$$\begin{aligned}\langle \mathbf{Su}, \mathbf{v} \rangle &= \langle \mathbf{u}, \mathbf{Sv} \rangle\\ \langle \mathbf{Fu}, \mathbf{v} \rangle &= -\langle \mathbf{u}, \mathbf{Fv} \rangle\end{aligned} \tag{6}$$

We evaluate Eqn. (3) by evaluating each term of L, given in Eqn. (4), except **V** which is trivially self-adjoint, and we get for the term $\mathbf{S}\,\partial_x^2$

$$\int_{-\infty}^{\infty} \langle \mathbf{Su''}, \mathbf{v} \rangle - \langle \mathbf{u}, \mathbf{Sv''} \rangle dx = \left[ \langle \mathbf{Su'}, \mathbf{v} \rangle - \langle \mathbf{u}, \mathbf{Sv'} \rangle \right]_{0+}^{0-} \qquad (7)$$

and for the term $\mathbf{F}\, \partial_x$

$$\int_{-\infty}^{\infty} \langle \mathbf{Fu'}, \mathbf{v} \rangle - \langle \mathbf{u}, \mathbf{Fv'} \rangle dx = -\left[ \langle \mathbf{u}, \mathbf{Fv} \rangle \right]_{0+}^{0-} \qquad (8)$$

Eqn. (8) can be proven by observing that

$$\int_{-\infty}^{\infty} \langle \mathbf{u}, \mathbf{Fv'} \rangle dx = \int_{-\infty}^{\infty} \sum_{i,j} \overline{u}_i f_{ij} v_j' dx =$$
$$= \left[ \sum_{i,j} \overline{u}_i f_{ij} v_j \right]_{0+}^{0-} - \int_{-\infty}^{\infty} \sum_{i,j} \overline{u}_i' f_{ij} v_j dx = \left[ \langle \mathbf{u}, \mathbf{Fv} \rangle \right]_{0+}^{0-} - \int_{-\infty}^{\infty} \langle \mathbf{u'}, \mathbf{Fv} \rangle dx = \qquad (9)$$
$$= \left[ \langle \mathbf{u}, \mathbf{Fv} \rangle \right]_{0+}^{0-} + \int_{-\infty}^{\infty} \langle \mathbf{Fu'}, \mathbf{v} \rangle dx$$

and a similar argument establishes Eqn. (7). The requirement of a self-adjoint operator $\mathbf{L}$, i. e. Eqns. (3) and (4), is thus given by

$$< L\mathbf{u} \mid \mathbf{v} > - < \mathbf{u} \mid L\mathbf{v} > = \left[ \langle \mathbf{u'}, \mathbf{Sv} \rangle - \langle \mathbf{u}, \mathbf{Sv'} \rangle - \langle \mathbf{u}, \mathbf{Fv} \rangle \right]_{0+}^{0-} = 0 \qquad (10)$$

We here see that it is not sufficient that $L$ is symmetric to be self-adjoint, which is a common misconception [14]. It is also necessary that the boundary conditions are correct. In order to solve Eqn. (10) we write

$$\mathbf{u}_+ = \mathbf{A}\mathbf{u}_- + \mathbf{B}\mathbf{u}_-' \qquad (11a)$$
$$\mathbf{u}_+' = \mathbf{C}\mathbf{u}_- + \mathbf{D}\mathbf{u}_-' \qquad (11b)$$

where the subscripts indicates the limit taken from the right (+) or the left (-) of the interface. We get by substitution in Eqn. (10)

$$\langle \mathbf{C}\mathbf{u}_- + \mathbf{D}\mathbf{u}_-', \mathbf{S}_+\mathbf{A}\mathbf{v}_- + \mathbf{S}_+\mathbf{B}\mathbf{v}_-' \rangle - \langle \mathbf{A}\mathbf{u}_- + \mathbf{B}\mathbf{u}_-', \mathbf{S}_+\mathbf{C}\mathbf{v}_- + \mathbf{S}_+\mathbf{D}\mathbf{v}_-' \rangle -$$
$$-\langle \mathbf{A}\mathbf{u}_- + \mathbf{B}\mathbf{u}_-', \mathbf{F}_+\mathbf{A}\mathbf{v}_- + \mathbf{F}_+\mathbf{B}\mathbf{v}_-' \rangle = \langle \mathbf{u}_-', \mathbf{S}_-\mathbf{v}_- \rangle - \langle \mathbf{u}_-, \mathbf{S}_-\mathbf{v}_-' \rangle - \langle \mathbf{u}_-, \mathbf{F}_-\mathbf{v}_- \rangle \qquad (12)$$

which is satisfied if

$$\mathbf{C}^*\mathbf{S}_+\mathbf{A} - \mathbf{A}^*\mathbf{S}_+\mathbf{C} - \mathbf{A}^*\mathbf{F}_+\mathbf{A} = -\mathbf{F}_-$$
$$\mathbf{D}^*\mathbf{S}_+\mathbf{B} - \mathbf{B}^*\mathbf{S}_+\mathbf{D} - \mathbf{B}^*\mathbf{F}_+\mathbf{B} = 0 \qquad (13)$$
$$\mathbf{D}^*\mathbf{S}_+\mathbf{A} - \mathbf{B}^*\mathbf{S}_+\mathbf{C} - \mathbf{B}^*\mathbf{F}_+\mathbf{A} = \mathbf{S}_-$$

By solving this equation set, one will get all allowed self-adjoint extensions. If we for simplicity assume **B**= **0** we have

$$\mathbf{u}_+ = \mathbf{A}\mathbf{u}_- \tag{14a}$$

$$\mathbf{u}_+' = \mathbf{C}\mathbf{u}_- + \mathbf{D}\mathbf{u}_-' \tag{14b}$$

which gives the simpler equation system

$$\mathbf{C}^*\mathbf{S}_+\mathbf{A} - \mathbf{A}^*\mathbf{S}_+\mathbf{C} - \mathbf{A}^*\mathbf{F}_+\mathbf{A} = -\mathbf{F}_- \tag{15a}$$

$$\mathbf{D}^*\mathbf{S}_+\mathbf{A} = \mathbf{S}_- \tag{15b}$$

We here see that the equations determining **A**, **C** and **D** are somewhat different to the single-band case [2,3] and (15a) has no analog in the single-band case. This equation arises from the presence of first derivatives in $L$ in the multi-band case. If **A** is a unit matrix we recover the standard boundary conditions with a continuous wavefunction as given e. g. in Ref. [15].

### III. The three-dimensional case

The three-dimensional case is considerably more involved than the one-dimensional case, although similar in principle. We first expand **L** as follows

$$\begin{aligned}\mathbf{L} = \mathbf{V} &+ \mathbf{F_x}\,\partial_x + \mathbf{F_y}\,\partial_y + \mathbf{F_z}\,\partial_z + \\ &+ \mathbf{S_{xx}}\,\partial_x^2 + \mathbf{S_{yy}}\,\partial_y^2 + \mathbf{S_{zz}}\,\partial_z^2 + \\ &+ \mathbf{S_{xy}}\,\partial_{xy} + \mathbf{S_{yx}}\,\partial_{yx} + \mathbf{S_{yz}}\,\partial_{yz} + \mathbf{S_{zy}}\,\partial_{zy} + \mathbf{S_{zx}}\,\partial_{zx} + \mathbf{S_{xz}}\,\partial_{xz}.\end{aligned} \tag{16}$$

The matrices **V**, $\mathbf{F}_i$ and $\mathbf{S}_{ij}$ do not contain derivatives and $\mathbf{S}_{ij} = \mathbf{S}_{ji}$ (which makes the expansion unique). After rearranging we get

$$\begin{aligned}\mathbf{L} &= \mathbf{V} + \mathbf{F_x}\,\partial_x + \mathbf{F_y}\,\partial_y + \mathbf{F_z}\,\partial_z + \mathbf{S_{xx}}\,\partial_x^2 + \mathbf{S_{yy}}\,\partial_y^2 + \mathbf{S_{zz}}\,\partial_z^2 + \\ &+ (\mathbf{S_{xy}}\,\partial_y + \mathbf{S_{xz}}\,\partial_z)\partial_x + (\mathbf{S_{yx}}\,\partial_x + \mathbf{S_{yz}}\,\partial_z)\partial_y + (\mathbf{S_{zx}}\,\partial_x + \mathbf{S_{zy}}\,\partial_y)\partial_z = \\ &= \mathbf{V} + \nabla\cdot\overline{\mathbf{F}} + \nabla\cdot\overline{\mathbf{T}}\end{aligned} \tag{17}$$

which also defines $\overline{\mathbf{F}}$ and $\overline{\mathbf{T}}$. Explicitly $\overline{\mathbf{T}} = (\mathbf{T_x}, \mathbf{T_y}, \mathbf{T_z})$ where

$$\mathbf{T_x} = \mathbf{S_{xx}}\,\partial_x + \mathbf{S_{xy}}\,\partial_y + \mathbf{S_{xz}}\,\partial_z = \nabla\cdot(\mathbf{S_{xx}}, \mathbf{S_{xy}}, \mathbf{S_{xz}}) = \nabla\cdot\overline{\mathbf{S}}_x \tag{18a}$$

$$\mathbf{T_y} = \mathbf{S_{yy}}\,\partial_y + \mathbf{S_{yx}}\,\partial_x + \mathbf{S_{yz}}\,\partial_z = \nabla\cdot(\mathbf{S_{yx}}, \mathbf{S_{yy}}, \mathbf{S_{yz}}) = \nabla\cdot\overline{\mathbf{S}}_y \tag{18b}$$

$$T_Z = S_{ZZ}\partial_z + S_{ZX}\partial_x + S_{ZY}\partial_y = \nabla \cdot (S_{ZX}, S_{ZY}, S_{ZZ}) = \nabla \cdot \overline{S}_z \tag{18c}$$

which defines $\overline{S}_i$ as well.
The adjoint matrices are given by

$$\mathbf{V}^* = \mathbf{V} \tag{19a}$$
$$\mathbf{F_i}^* = -\mathbf{F_i} \tag{19b}$$
$$\mathbf{S_{ij}}^* = \mathbf{S_{ji}} = \mathbf{S_{ij}} \tag{19c}$$

In order to simplify the notation we define "scalar products" of several vector-valued functions as

$$\langle (\mathbf{u}_1, \mathbf{u}_2, \mathbf{u}_3), (\mathbf{v}_1, \mathbf{v}_2, \mathbf{v}_3) \rangle = \langle \mathbf{u}_1, \mathbf{v}_1 \rangle + \langle \mathbf{u}_2, \mathbf{v}_2 \rangle + \langle \mathbf{u}_3, \mathbf{v}_3 \rangle \tag{20a}$$
$$\langle \mathbf{u}_1, (\mathbf{v}_1, \mathbf{v}_2, \mathbf{v}_3) \rangle = (\langle \mathbf{u}_1, \mathbf{v}_1 \rangle, \langle \mathbf{u}_2, \mathbf{v}_2 \rangle, \langle \mathbf{u}_3, \mathbf{v}_3 \rangle) \tag{20b}$$

A definition and two useful relations, used in the rest of this section are

$$(\mathbf{F_X}, \mathbf{F_Y}, \mathbf{F_Z})\mathbf{u} = (\mathbf{F_X u}, \mathbf{F_Y u}, \mathbf{F_Z u}) \tag{21}$$
$$(\nabla \cdot (\mathbf{F_X}, \mathbf{F_Y}, \mathbf{F_Z}))\mathbf{u} = \nabla \cdot ((\mathbf{F_X}, \mathbf{F_Y}, \mathbf{F_Z})\mathbf{u}) \tag{22}$$
$$\nabla \cdot \overline{\mathbf{F}} = \overline{\mathbf{F}} \cdot \nabla \tag{23}$$

and analogously for the other operators ($\overline{\mathbf{T}}$ and $\overline{\mathbf{S}}_i$).
The three-dimensional analogue of Eqn. (3) becomes

$$\int_V \langle L\mathbf{u}, \mathbf{v}\rangle - \langle \mathbf{u}, L\mathbf{v}\rangle dr^3 = \int_{V_1} \langle L\mathbf{u}, \mathbf{v}\rangle - \langle \mathbf{u}, L\mathbf{v}\rangle dr^3 + \int_{V_\infty} \langle L\mathbf{u}, \mathbf{v}\rangle - \langle \mathbf{u}, L\mathbf{v}\rangle dr^3 \tag{24}$$

where all space, V, is divided into a closed volume $V_1$ and a volume extending to infinity, $V_\infty$. The discontinuities in the coefficients of $L$ occur at the boundary of $V_1$. We will evaluate Eqn. (24) by evaluating each term of $L$ given in Eqn. (17) except $\mathbf{V}$ which is trivially self-adjoint. We get by substituting $\nabla \cdot \overline{\mathbf{F}}$ for $L$ in the integral over $V_1$ in Eqn. (24)

$$\int_{V_1} \langle \nabla \cdot \overline{\mathbf{F}}\mathbf{u}, \mathbf{v}\rangle - \langle \mathbf{u}, \nabla \cdot \overline{\mathbf{F}}\mathbf{v}\rangle dr^3 =$$

$$= \int_{V_1} \langle \overline{\mathbf{F}} \cdot \nabla \mathbf{u}, \mathbf{v}\rangle - \langle \mathbf{u}, \nabla \cdot \overline{\mathbf{F}}\mathbf{v}\rangle dr^3 = \int_{V_1} -\langle \nabla \mathbf{u}, \overline{\mathbf{F}}\mathbf{v}\rangle - \langle \mathbf{u}, \nabla \cdot \overline{\mathbf{F}}\mathbf{v}\rangle dr^3 = \tag{25}$$

$$= -\int_{V_1} \nabla \cdot \langle \mathbf{u}, \overline{\mathbf{F}}\mathbf{v}\rangle dr^3 = -\int_{\partial V_{1-}} \langle \mathbf{u}, \overline{\mathbf{F}}\mathbf{v}\rangle \cdot \mathbf{n} dS$$

by Eqns. (19-21) and Gauss theorem. The last integral is a surface integral and by $\partial V_{1-}$ ($\partial V_{1+}$) we mean the boundary of $V_1$ and the integrand takes its limits from the inside (outside) of $V_1$. By applying the same argument to the integral over $V_\infty$ we get

$$\int_{V_\infty} \langle \nabla \cdot \mathbf{u}, \overline{\mathbf{F}}\mathbf{v}\rangle - \langle \mathbf{u}, \nabla \cdot \overline{\mathbf{F}}\mathbf{v}\rangle dr^3 = \int_{\partial V_{1+}} \langle \mathbf{u}, \overline{\mathbf{F}}\mathbf{v}\rangle \cdot \mathbf{n} dS \qquad (26)$$

and thus

$$\int_V \langle \nabla \cdot \mathbf{u}, \overline{\mathbf{F}}\mathbf{v}\rangle - \langle \mathbf{u}, \nabla \cdot \overline{\mathbf{F}}\mathbf{v}\rangle dr^3 = (\int_{\partial V_{1+}} - \int_{\partial V_{1-}}) \langle \mathbf{u}, \overline{\mathbf{F}}\mathbf{v}\rangle \cdot \mathbf{n} dS \qquad (27)$$

The situation for the third term, $\nabla \cdot \overline{\mathbf{T}}$ in Eqn. (17), involving second and mixed derivatives is more complicated. We will initially essentially state the results in Eqn. (28) and (30) below and then supply a proof. We have

$$\int_{V_1} \langle \nabla \cdot \overline{\mathbf{T}}\mathbf{u}, \mathbf{v}\rangle - \langle \mathbf{u}, \nabla \cdot \overline{\mathbf{T}}\mathbf{v}\rangle dr^3 =$$

$$\int_{V_1} \langle \nabla \cdot (\mathbf{T}_x, \mathbf{T}_y, \mathbf{T}_z)\mathbf{u}, \mathbf{v}\rangle - \langle \mathbf{u}, \nabla \cdot (\mathbf{T}_x, \mathbf{T}_y, \mathbf{T}_z)\mathbf{v}\rangle dr^3 =$$

$$= \int_{V_1} \nabla \cdot \langle (\mathbf{T}_x, \mathbf{T}_y, \mathbf{T}_z)\mathbf{u}, \mathbf{v}\rangle - \langle \mathbf{u}, (\mathbf{T}_x, \mathbf{T}_y, \mathbf{T}_z)\mathbf{v}\rangle\rangle dr^3 = \qquad (28)$$

$$\int_{\partial V_-} \langle (\mathbf{T}_x, \mathbf{T}_y, \mathbf{T}_z)\mathbf{u}, \mathbf{v}\rangle - \langle \mathbf{u}, (\mathbf{T}_x, \mathbf{T}_y, \mathbf{T}_z)\mathbf{v}\rangle\rangle \cdot \mathbf{n} dS$$

by Gauss theorem. For the integral over $\mathbf{V}_\infty$ we analogously get

$$\int_{V_\infty} \langle \nabla \cdot \overline{\mathbf{T}}\mathbf{u}, \mathbf{v}\rangle - \langle \mathbf{u}, \nabla \cdot \overline{\mathbf{T}}\mathbf{v}\rangle dr^3 = -\int_{\partial V_+} \langle (\mathbf{T}_x, \mathbf{T}_y, \mathbf{T}_z)\mathbf{u}, \mathbf{v}\rangle - \langle \mathbf{u}, (\mathbf{T}_x, \mathbf{T}_y, \mathbf{T}_z)\mathbf{v}\rangle\rangle \cdot \mathbf{n} dS \qquad (29)$$

and we thus have for the integral over all space

$$\int_V \langle \nabla \cdot \overline{\mathbf{T}}\mathbf{u}, \mathbf{v}\rangle - \langle \mathbf{u}, \nabla \cdot \overline{\mathbf{T}}\mathbf{v}\rangle dr^3 =$$

$$(\int_{\partial V_-} - \int_{\partial V_+}) \langle (\mathbf{T}_x, \mathbf{T}_y, \mathbf{T}_z)\mathbf{u}, \mathbf{v}\rangle - \langle \mathbf{u}, (\mathbf{T}_x, \mathbf{T}_y, \mathbf{T}_z)\mathbf{v}\rangle\rangle \cdot \mathbf{n} dS \qquad (30)$$

Equation (28) is not trivial and we will prove it. The first equality in (28) is by definition. In order to prove the second equality in Eqn. (28) we expand

$$\nabla \cdot \langle (\mathbf{T}_x, \mathbf{T}_y, \mathbf{T}_z)\mathbf{u}, \mathbf{v}\rangle - \langle \mathbf{u}, (\mathbf{T}_x, \mathbf{T}_y, \mathbf{T}_z)\mathbf{v}\rangle) =$$

$$= \nabla \cdot (((\langle \mathbf{T}_x\mathbf{u}, \mathbf{v}\rangle, \langle \mathbf{T}_y\mathbf{u}, \mathbf{v}\rangle, \langle \mathbf{T}_z\mathbf{u}, \mathbf{v}\rangle) - (\langle \mathbf{u}, \mathbf{T}_x\mathbf{v}\rangle, \langle \mathbf{u}, \mathbf{T}_y\mathbf{v}\rangle, \langle \mathbf{u}, \mathbf{T}_z\mathbf{v}\rangle)) \qquad (31)$$

$$= (\partial_x \langle \mathbf{T}_x\mathbf{u}, \mathbf{v}\rangle + \partial_y \langle \mathbf{T}_y\mathbf{u}, \mathbf{v}\rangle + \partial_x \langle \mathbf{T}_z\mathbf{u}, \mathbf{v}\rangle) - (\partial_x \langle \mathbf{u}, \mathbf{T}_x\mathbf{v}\rangle + \partial_y \langle \mathbf{u}, \mathbf{T}_y\mathbf{v}\rangle + \partial_x \langle \mathbf{u}, \mathbf{T}_z\mathbf{v}\rangle)$$

If we look at the term involving $\partial_x$ we have

$$\partial_x \langle \mathbf{T}_x \mathbf{u}, \mathbf{v} \rangle - \partial_x \langle \mathbf{u}, \mathbf{T}_x \mathbf{v} \rangle = \langle \mathbf{T}_x \partial_x \mathbf{u}, \mathbf{v} \rangle + \langle \mathbf{T}_x \mathbf{u}, \partial_x \mathbf{v} \rangle - \langle \partial_x \mathbf{u}, \mathbf{T}_x \mathbf{v} \rangle - \langle \mathbf{u}, \mathbf{T}_x \partial_x \mathbf{v} \rangle. \tag{32}$$

It is thus clear that if we can show that

$$\sum_{i=x,y,z} \langle \mathbf{T}_i \mathbf{u}, \partial_i \mathbf{v} \rangle - \langle \partial_i \mathbf{u}, \mathbf{T}_i \mathbf{v} \rangle = 0 \tag{33}$$

we have

$$\nabla \cdot \langle (\mathbf{T}_x, \mathbf{T}_y, \mathbf{T}_z) \mathbf{u}, \mathbf{v} \rangle - \langle \mathbf{u}, (\mathbf{T}_x, \mathbf{T}_y, \mathbf{T}_z) \mathbf{v} \rangle \rangle =$$
$$= (\partial_x \langle \mathbf{T}_x \mathbf{u}, \mathbf{v} \rangle + \partial_y \langle \mathbf{T}_y \mathbf{u}, \mathbf{v} \rangle + \partial_z \langle \mathbf{T}_z \mathbf{u}, \mathbf{v} \rangle) - (\partial_x \langle \mathbf{u}, \mathbf{T}_x \mathbf{v} \rangle + \partial_y \langle \mathbf{u}, \mathbf{T}_y \mathbf{v} \rangle + \partial_x \langle \mathbf{u}, \mathbf{T}_z \mathbf{v} \rangle) = \tag{34}$$
$$= \sum_{i=x,y,z} \langle \mathbf{T}_i \partial_i \mathbf{u}, \mathbf{v} \rangle - \langle \mathbf{u}, \mathbf{T}_i \partial_i \mathbf{v} \rangle = \langle \nabla \cdot (\mathbf{T}_x, \mathbf{T}_y, \mathbf{T}_z) \mathbf{u}, \mathbf{v} \rangle - \langle \mathbf{u}, \nabla \cdot (\mathbf{T}_x, \mathbf{T}_y, \mathbf{T}_z) \mathbf{v} \rangle$$

which is the desired result. We evaluate Eqn. (33)

$$\sum_{i=x,y,z} \langle \mathbf{T}_i \mathbf{u}, \partial_i \mathbf{v} \rangle - \langle \partial_i \mathbf{u}, \mathbf{T}_i \mathbf{v} \rangle =$$
$$\langle \mathbf{S}_{xy} \partial_y \mathbf{u}, \partial_x \mathbf{v} \rangle + \langle \mathbf{S}_{xz} \partial_z \mathbf{u}, \partial_x \mathbf{v} \rangle - \langle \partial_x \mathbf{u}, \mathbf{S}_{xy} \partial_y \mathbf{v} \rangle - \langle \partial_x \mathbf{u}, \mathbf{S}_{xz} \partial_z \mathbf{v} \rangle +$$
$$\langle \mathbf{S}_{yz} \partial_z \mathbf{u}, \partial_y \mathbf{v} \rangle + \langle \mathbf{S}_{yx} \partial_x \mathbf{u}, \partial_y \mathbf{v} \rangle - \langle \partial_y \mathbf{u}, \mathbf{S}_{yz} \partial_z \mathbf{v} \rangle - \langle \partial_y \mathbf{u}, \mathbf{S}_{yx} \partial_x \mathbf{v} \rangle + \tag{35}$$
$$\langle \mathbf{S}_{zx} \partial_x \mathbf{u}, \partial_z \mathbf{v} \rangle + \langle \mathbf{S}_{zy} \partial_y \mathbf{u}, \partial_z \mathbf{v} \rangle - \langle \partial_z \mathbf{u}, \mathbf{S}_{zx} \partial_x \mathbf{v} \rangle - \langle \partial_z \mathbf{u}, \mathbf{S}_{zy} \partial_y \mathbf{v} \rangle +$$
$$\langle \mathbf{S}_{xx} \partial_x \mathbf{u}, \partial_x \mathbf{v} \rangle - \langle \partial_x \mathbf{u}, \mathbf{S}_{xx} \partial_x \mathbf{v} \rangle + \langle \mathbf{S}_{yy} \partial_y \mathbf{u}, \partial_y \mathbf{v} \rangle - \langle \partial_y \mathbf{u}, \mathbf{S}_{yy} \partial_y \mathbf{v} \rangle +$$
$$\langle \mathbf{S}_{zz} \partial_z \mathbf{u}, \partial_z \mathbf{v} \rangle - \langle \partial_z \mathbf{u}, \mathbf{S}_{zz} \partial_z \mathbf{v} \rangle$$

The right hand side of Eqn. (35) is zero since $\mathbf{S}_{ij} = \mathbf{S}_{ji}*$, by Eqn. (19c).
Thus Eqns. (33) and (28) and consequently (30) are true. Summarising, the condition for *L* to be selfadjoint is

$$\int_V \langle L\mathbf{u}, \mathbf{v} \rangle - \langle \mathbf{u}, L\mathbf{v} \rangle dr^3 =$$
$$(\int_{\partial V_-} - \int_{\partial V_+}) (\langle (\mathbf{T}_x, \mathbf{T}_y, \mathbf{T}_z) \mathbf{u}, \mathbf{v} \rangle - \langle \mathbf{u}, (\mathbf{T}_x, \mathbf{T}_y, \mathbf{T}_z) \mathbf{v} \rangle - \langle \mathbf{u}, \overline{\mathbf{F}} \mathbf{v} \rangle) \cdot \mathbf{n} dS = 0 \tag{36}$$

In the one-dimensional case we found that the boundary conditions could be parametrised with a finite number of parameters (implicitly given by Eqns. 13 or 15). In the three-dimensional case the parameter space is not even finite in dimension. By imposing suitable constraints it is possible to find families of boundary conditions which depend on the curvature (Gaussian or mean) [13] of the interface. We mention that if a wave-function is restricted to a two-dimensional manifold, there exist extra terms in the Hamiltonian arising from the curvature of the manifold [17]. If we require that the boundary conditions are local to each point in the interface, which seems more natural, we find that

$$(\langle (\mathbf{T}_{x+},\mathbf{T}_{y+},\mathbf{T}_{z+})\mathbf{u}_{+},\mathbf{v}_{+}\rangle - \langle \mathbf{u}_{+},(\mathbf{T}_{x+},\mathbf{T}_{y+},\mathbf{T}_{z+})\mathbf{v}_{+}\rangle - \langle \mathbf{u}_{+},\overline{\mathbf{F}}_{+}\mathbf{v}_{+}\rangle) \cdot \mathbf{n} = \tag{37}$$
$$(\langle (\mathbf{T}_{x-},\mathbf{T}_{y-},\mathbf{T}_{z-})\mathbf{u}_{-},\mathbf{v}_{-}\rangle - \langle \mathbf{u}_{-},(\mathbf{T}_{x-},\mathbf{T}_{y-},\mathbf{T}_{z-})\mathbf{v}_{-}\rangle - \langle \mathbf{u}_{-},\overline{\mathbf{F}}_{-}\mathbf{v}_{-}\rangle) \cdot \mathbf{n}$$

which is similar to Eqn. (10) which gives the one-dimensional case. A useful consistency check is obtained by using $\mathbf{n} = (100)$ in the equation above, corresponding to an interface normal to the x-direction and which should give equivalent results as in the one-dimensional case. We then get

$$\langle \mathbf{T}_{x+}\mathbf{u}_{+},\mathbf{v}_{+}\rangle - \langle \mathbf{u}_{+},\mathbf{T}_{x+}\mathbf{v}_{+}\rangle - \langle \mathbf{u}_{+},\mathbf{F}_{x+}\mathbf{v}_{+}\rangle = \langle \mathbf{T}_{x-}\mathbf{u}_{-},\mathbf{v}_{-}\rangle - \langle \mathbf{u}_{-},\mathbf{T}_{x-}\mathbf{v}_{-}\rangle - \langle \mathbf{u}_{-},\mathbf{F}_{x-}\mathbf{v}_{-}\rangle \tag{38}$$

or equivalently by Eqn. (18a)

$$\langle \nabla \mathbf{u}_{+},\overline{\mathbf{S}}_{x+}\mathbf{v}_{+}\rangle - \langle \mathbf{u}_{+},\overline{\mathbf{S}}_{x+}\nabla\mathbf{v}_{+}\rangle - \langle \mathbf{u}_{+},\mathbf{F}_{x+}\mathbf{v}_{+}\rangle = \langle \nabla \mathbf{u}_{-},\overline{\mathbf{S}}_{x-}\mathbf{v}_{-}\rangle - \langle \mathbf{u}_{-},\overline{\mathbf{S}}_{x-}\nabla\mathbf{v}_{-}\rangle - \langle \mathbf{u}_{-},\mathbf{F}_{x-}\mathbf{v}_{-}\rangle \tag{39}$$

and then

$$\begin{aligned}
&\langle \partial_x\mathbf{u}_{+},\mathbf{S}_{xx+}\mathbf{v}_{+}\rangle + \langle \partial_y\mathbf{u}_{+},\mathbf{S}_{xy+}\mathbf{v}_{+}\rangle + \langle \partial_z\mathbf{u}_{+},\mathbf{S}_{xz+}\mathbf{v}_{+}\rangle - \\
&-\langle \mathbf{u}_{+},\mathbf{S}_{xx+}\partial_x\mathbf{v}_{+}\rangle - \langle \mathbf{u}_{+},\mathbf{S}_{xy+}\partial_y\mathbf{v}_{+}\rangle - \langle \mathbf{u}_{+},\mathbf{S}_{xz+}\partial_z\mathbf{v}_{+}\rangle - \langle \mathbf{u}_{+},\mathbf{F}_{x+}\mathbf{v}_{+}\rangle = \\
&\langle \partial_x\mathbf{u}_{-},\mathbf{S}_{xx-}\mathbf{v}_{-}\rangle + \langle \partial_y\mathbf{u}_{-},\mathbf{S}_{xy-}\mathbf{v}_{-}\rangle + \langle \partial_z\mathbf{u}_{-},\mathbf{S}_{xz-}\mathbf{v}_{-}\rangle - \\
&-\langle \mathbf{u}_{-},\mathbf{S}_{xx-}\partial_x\mathbf{v}_{-}\rangle - \langle \mathbf{u}_{-},\mathbf{S}_{xy-}\partial_y\mathbf{v}_{-}\rangle - \langle \mathbf{u}_{-},\mathbf{S}_{xz-}\partial_z\mathbf{v}_{-}\rangle - \langle \mathbf{u}_{-},\mathbf{F}_{x-}\mathbf{v}_{-}\rangle
\end{aligned} \tag{40}$$

Now since $\partial_y \mathbf{u} = ik_y\mathbf{u}$ and $\partial_z\mathbf{u} = ik_z\mathbf{u}$ (and similarly for $\mathbf{v}$) in the one-dimensional case, we finally get

$$\begin{aligned}
&\langle \partial_x\mathbf{u}_{+},\mathbf{S}_{xx+}\mathbf{v}_{+}\rangle - \langle \mathbf{u}_{+},\mathbf{S}_{xx+}\partial_x\mathbf{v}_{+}\rangle - \langle \mathbf{u}_{+},\mathbf{F}_{+}\mathbf{v}_{+}\rangle = \\
&= \langle \partial_x\mathbf{u}_{-},\mathbf{S}_{xx-}\mathbf{v}_{-}\rangle - \langle \mathbf{u}_{-},\mathbf{S}_{xx-}\partial_x\mathbf{v}_{-}\rangle - \langle \mathbf{u}_{-},\mathbf{F}_{-}\mathbf{v}_{-}\rangle
\end{aligned} \tag{41}$$

which agrees with one-dimensional case, Eqn. (10). Note that

$$\mathbf{F}=\mathbf{F_x}-ik_yS_{xy}-ik_zS_{xz}. \tag{42}$$

The choice of boundary conditions will affect the energy levels, and for example, may introduce gap-states [18,19].

## IV. Numerical calculations

The discussion in section II and III gives a mathematical treatment of the choice of permissible boundary conditions. It is of interest to check numerically how much different boundary conditions affect the energy levels in e. g. quantum wells. One approach is to explicitely match the wave-functions in the barrier with the wave-functions in the well as explained by Eppenga et al [15]. This approach does not generalise to quantum wires and dots, however, and we have instead used a method which is very similar to a plane-wave expansion. In order to impart the correct behaviour at the interfaces we use a basis set such that each wavefunction (in the base) has the chosen boundary conditions. The wavefunctions will typically not be orthogonal, although they form a complete set, and a generalised eigenvalue problem has to be solved. Mixing among eight bands is included in the model.

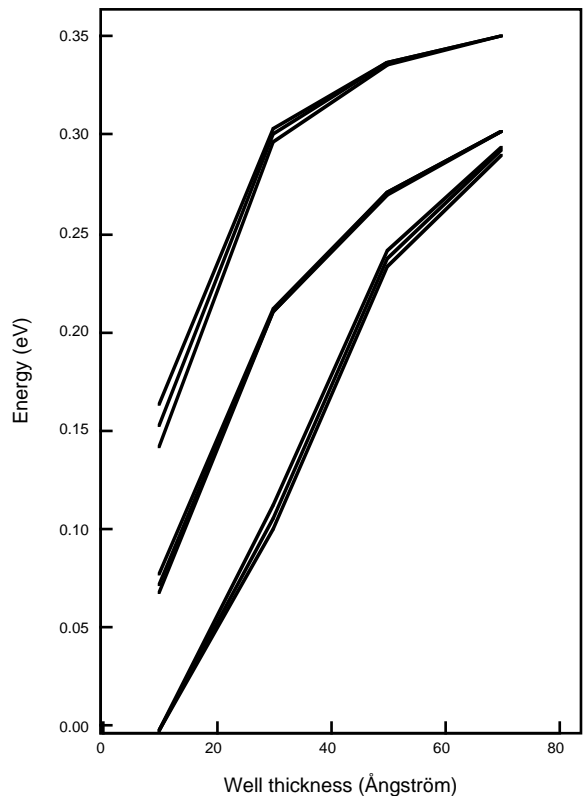

*Figure 1. A plot of the hole energy levels in an InGaAs quantum well, inbetween barriers of InP, as a function of the well thickness. The zero-level corresponds to the top of the valence band in the barrier (InP). Three states have been calculated and for each state the confinement energy is highest (and the absolute energy lowest) for a small value of j (defined in the text).*

We have checked our computer program against the results of Gershoni et al. [20], using the standard boundary conditions, and find complete agreement. We have chosen to model an InGaAs quantum well, lattice matched to InP.

The materials parameters were taken from Ref. [20]. As boundary conditions we use Eqns. 14a and b, where **A** is taken to be a unit matrix times a factor (j). Thus the wavefunctions will thus have jumps at the interfaces, which we consider to be the most interesting case. In Figure 1 we show the energy levels in the valence band as a function of the thickness of the quantum well. Three different values of j have been used: 0.9, 1.0 and 1.1. If j is larger than one there is a jump such that the wavefunction in the well has a higher amplitude than if continuous wavefunctions are used (j = 1). We find that the confinement energy is smaller for larger values of j. This effect is most pronounced for the ground state and amounts to about 10 meV. The physical reason is that a large value of j will increase the amplitude of the wavefunction in the well and the barrier will have less effect, thus decreasing the confinement energy. In fact, for j=0 there will be no effect of the well at all, since the amplitude of the wavefunction is then zero in the well.

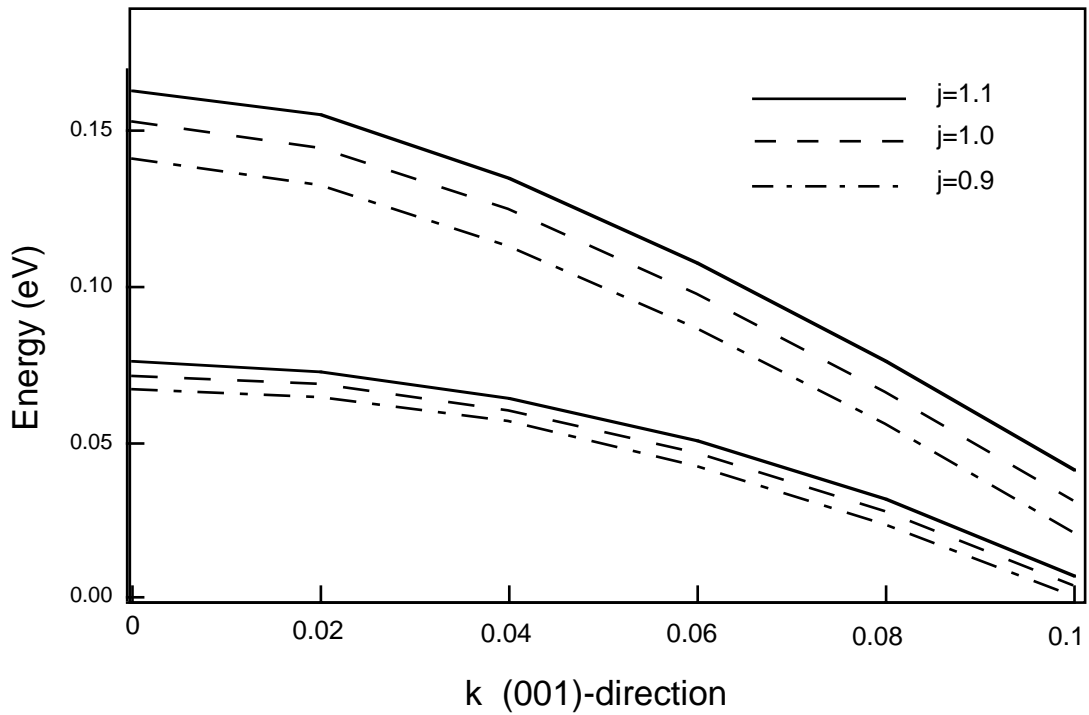

*Figure 2. A dispersion diagram for a quantum well of InGaAs in InP. The well thickness is 10 Å. Tree different values of the jump-coefficent j, explained in the main text, have been used.*

In Figure 2 we show the energy versus k-values for a 10 Ångström thick quantum well. We find that the energy shift is almost independent of k.

**Acknowledgement:** This work was performed within the nanometer structure consortium in Lund, and was supported by NFR, TFR, SSF, and Nutek.

**Appendix**:

The Kane-matrix, using the basis $|s\uparrow\rangle$, $|x\uparrow\rangle$, $|y\uparrow\rangle$, $|z\downarrow\rangle$, $|s\downarrow\rangle$, $|x\downarrow\rangle$, $|y\downarrow\rangle$, and $|z\downarrow\rangle$, is given by [8,13]:

$$H = \begin{bmatrix} G(\mathbf{k}) & \Gamma \\ -\Gamma^* & G^*(-\mathbf{k}) \end{bmatrix} \tag{A1}$$

where $G = G_1 + G_2 + G_{so}$ and $\Gamma$ are given by

$$G_1 = \begin{bmatrix} E_c & iPk_x & iPk_y & iPk_z \\ -iPk_x & E_v & 0 & 0 \\ -iPk_y & 0 & E_v & 0 \\ -iPk_z & 0 & 0 & E_v \end{bmatrix}, \tag{A2}$$

$$G_2 = \begin{bmatrix} A'k^2 & Bk_y k_z & Bk_x k_z & Bk_x k_y \\ Bk_y k_z & L'k_x^2 + M(k_y^2 + k_z^2) & Nk_x k_y & Nk_x k_z \\ Bk_x k_z & Nk_x k_y & L'k_y^2 + M(k_x^2 + k_z^2) & Nk_y k_z \\ Bk_x k_y & Nk_x k_z & Nk_y k_z & L'k_z^2 + M(k_x^2 + k_y^2) \end{bmatrix}, \tag{A3}$$

$$G_{so} = -\frac{\Delta}{3}\begin{bmatrix} 0 & 0 & 0 & 0 \\ 0 & 0 & i & 0 \\ 0 & -i & 0 & 0 \\ 0 & 0 & 0 & 0 \end{bmatrix}, \tag{A4}$$

$$\Gamma = -\frac{\Delta}{3}\begin{bmatrix} 0 & 0 & 0 & 0 \\ 0 & 0 & 0 & -1 \\ 0 & 0 & 0 & i \\ 0 & 1 & -i & 0 \end{bmatrix} \tag{A5}$$

where all parameters (A', B, ... ) are real numbers. The matrix H is converted to a differential operator by the replacement, $k_t \to \partial_t/i$. The matrices which are used in the main text can easily be obtained from these expressions.